\documentclass[pdflatex,sn-mathphys-num]{sn-jnl}
\usepackage[T1]{fontenc}%
\usepackage{graphicx}%
\usepackage{multirow}%
\usepackage{amsmath,amssymb,amsfonts}%
\usepackage{amsthm}%
\usepackage{mathrsfs}%
\usepackage[title]{appendix}%
\usepackage{textcomp}%
\usepackage{manyfoot}%
\usepackage{booktabs}%
\usepackage[normalem]{ulem}
\usepackage[ruled, lined, linesnumbered, commentsnumbered, longend]{algorithm2e}
\usepackage{algpseudocode}%
\usepackage{listings}%
\usepackage{hyperref}%
\usepackage{makecell}%
\usepackage{siunitx}
\usepackage{algpseudocode}
\usepackage{graphicx}
\usepackage{anyfontsize}
\usepackage{geometry}
\usepackage{booktabs}
\usepackage{tabularx}
\usepackage{multicol}
\usepackage{threeparttable}
\usepackage{ulem}
\usepackage{xcolor}
\definecolor{blue}{RGB}{0,80,200}
\usepackage[section]{placeins}
\theoremstyle{thmstyleone}%
\theoremstyle{thmstyletwo}%

\theoremstyle{thmstylethree}%

\begin{document}

\title[Article Title]{Pretraining a Foundation Model for Small-Molecule Natural Products}


\author[1]{\fnm{Yuheng} \sur{Ding}}\email{1910307414@pku.edu.cn}
\equalcont{These authors contributed equally to this work.}
\author[2]{\fnm{Bo} \sur{Qiang}}\email{bqiang@uw.edu}
\equalcont{These authors contributed equally to this work.}
\author[3]{\fnm{Shaoning} \sur{Li}}\email{snli24@cse.cuhk.edu.hk}
\author[1]{\fnm{Yiran} \sur{Zhou}}\email{yrzhou@bjmu.edu.cn}
\author[1]{\fnm{Jie} \sur{Yu}}\email{jieyu@bjmu.edu.cn}
\author[1]{\fnm{Qi} \sur{Li}}\email{liqi@stu.pku.edu.cn}
\author[1]{\fnm{Cheng} \sur{Shi}}\email{shicheng\_pku@163.com}
\author[1]{\fnm{Liangren} \sur{Zhang}}\email{liangren@bjmu.edu.cn}
\author*[4]{\fnm{Yusong} \sur{Wang}}\email{wangyusong2000@stu.xjtu.edu.cn}
\author*[4]{\fnm{Nanning} \sur{Zheng}}\email{nnzheng@xjtu.edu.cn}
\author*[1,5]{\fnm{Zhenming} \sur{Liu}}\email{zmliu@bjmu.edu.cn}

\affil*[1]{\orgdiv{State Key Laboratory of Natural and Biomimetic Drugs}, \orgname{School of Pharmaceutical Sciences, Peking University}, \orgaddress{\city{Beijing}, \postcode{100191},  \country{China}}}

\affil[2]{\orgdiv{Paul G. Allen School of Computer Science \& Engineering}, \orgname{University of Washington}, \orgaddress{\city{Seattle}, \postcode{98105}, \state{WA}, \country{US}}}

\affil[3]{\orgdiv{Department of Computer Science \& Engineering},\orgname{The Chinese University of Hong Kong},\orgaddress{\city{Hong Kong},\country{China}}}

\affil*[4]{\orgdiv{State Key Laboratory of Human-Machine Hybrid Augmented Intelligence}, \orgname{Institute of Artificial Intelligence and Robotics, Xi’an Jiaotong University}, \orgaddress{\city{Xi’an}, \country{China}}}

\affil*[5]{\orgdiv{Key Laboratory of Xinjiang Endemic Phytomedicine Resources Ministry of Education}, \orgname{School of Pharmacy, Shihezi University}, \orgaddress{\city{Shihezi, Xinjiang}, \postcode{832003}, \country{China}}}


\abstract{
Natural products, as metabolites from microorganisms, animals, or plants, exhibit diverse biological activities, making them crucial for drug discovery.
Nowadays, existing deep learning methods for natural products research primarily rely on supervised learning approaches designed for specific downstream tasks. However, such one-model-for-a-task paradigm often lacks generalizability and leaves significant room for performance improvement.
Additionally, existing molecular characterization methods are not well-suited for the unique tasks associated with natural products. To address these limitations, we have pre-trained a foundation model for natural products (NaFM) based on their unique properties.
Our approach employs a pre-training strategy specifically tailored to natural products. By incorporating contrastive learning and masked graph learning objectives, we emphasize evolutional information from molecular scaffolds while capturing side-chain information.
NaFM achieves state-of-the-art (SOTA) results in various downstream tasks related to natural product mining and drug discovery. 
We first compare taxonomy classification with synthesized molecule-focused baselines to demonstrate that current models are inadequate for understanding natural synthesis. 
Furthermore, by diving into a fine-grained analysis at both the gene and microbial levels, NaFM demonstrates the ability to capture evolutionary information. Eventually, our method is experimented with virtual screening, illustrating informative natural product representations that can lead to more effective identification of potential drug candidates.}

\maketitle
\section{Main}\label{sec1}

Natural product(NPs) are metabolites or secondary metabolites produced by bacteria, fungi, animals, plants, and other organisms. Their structural diversity is immense, encompassing hundreds of thousands of distinct compounds~\citep{chandrasekhar2025coconut}. As metabolites, they are inherently bioactive, making them valuable candidates for direct pharmaceutical development or structural modification into therapeutic agents—particularly for anticancer and anti-infective drugs~\citep{newman2016natural,clark1996natural,harvey2008natural}.
Despite their value, discovering natural compounds suitable for drug development remains time-consuming and costly~\citep{li2009drug}. Traditional bioactivity-guided discovery involves extracting metabolites using solvents of varying polarity, followed by successive bioactivity-guided fractionation until pure active compounds are isolated~\citep{atanasov2021natural}. This process often requires identifying both the structural class of the compound and the producing microorganism to avoid unnecessary cultivation. Moreover, phenotypic screening of bioactive compounds demands substantial effort to determine their molecular targets~\citep{corson2007molecular}.
General small-molecule databases such as ZINC~\citep{irwin2020zinc20} and Enamine REAL~\citep{EnamineREALDatabase} enable large-scale virtual screening; for instance, \citet{stokes2020deep} leveraged machine-learning models to identify novel antibiotics from vast structural libraries. Building on this foundation, the natural-products community has curated extensive domain-specific resources, including comprehensive compound collections~\citep{banerjee2015super,sorokina2021coconut,rutz2022lotus}, activity and biological-origin databases~\citep{zeng2018npass,van2019natural}, repositories of biosynthetic gene clusters (BGCs)~\citep{terlouw2023mibig}, and specialized marine natural-product datasets~\citep{lei2002marine,barbosa2019free,lyu2021cmnpd}. Concurrently, machine-learning approaches have been developed to mine these resources~\citep{aghdam2021deep,zheng2022deep,lai2020privileged,yoo2020deep,hannigan2019deep,liu2021deep}; for example, \citet{xu2024composite} applied deep learning to the taxonomic classification of NPs.
Commonly used natural-product classification tools such as ClassyFire~\citep{djoumbou2016classyfire} and NPClassifier~\citep{kim2021npclassifier} rely solely on supervised learning. However, the limited availability of labeled data and inherent dataset biases often reduce their accuracy and generalization, particularly for out-of-distribution or less-related downstream tasks. Despite these limitations, they remain standard classification tools across many natural-product databases.

Pre-training on large-scale unlabeled molecular data has emerged as an effective strategy for representation learning~\citep{yu2021review}. Early methods borrowed natural language processing techniques, encoding SMILES strings~\citep{weininger1989smiles} with RNNs~\citep{cho2014learning} or Transformers~\citep{vaswani2017attention} for masked language modeling~\citep{xu2017seq2seq, jastrzkebski2016learning}. Subsequent studies showed that graph neural networks~\citep{kearnes2016molecular, schutt2017schnet} and graph-specific pre-training strategies~\citep{wang2022molecular, hu2019strategies, xia2023mole} yield superior results, while integrating 2D graphs with 3D structures, semantic information, or force-field data further enhances performance~\citep{liu2021pre, zhu2022unified, li2023knowledge, ni2024pre}.

Despite these advances, few foundation models target NPs, which differ fundamentally from synthetic molecules in structural diversity and complexity~\citep{mullowney2023artificial}. Their unique scaffolds, stereochemistry, and functional groups challenge existing 3D or force-field–based approaches. Yet, NPs exhibit a distinctive organizing principle: analogous to the central dogma of molecular biology, intrinsic relationships link their biological origins, scaffold architectures, and biochemical properties.
Similar compounds are more likely to originate from related biological sources that share overlapping biosynthetic machinery. In particular, NPs with comparable scaffolds or functional motifs often arise from similar genes or BGCs, which encode enzymes or protein substrates that direct specific metabolic pathways leading to conserved structural frameworks.

\begin{figure*}[htbp]
    \centering
    \includegraphics[width=\textwidth]{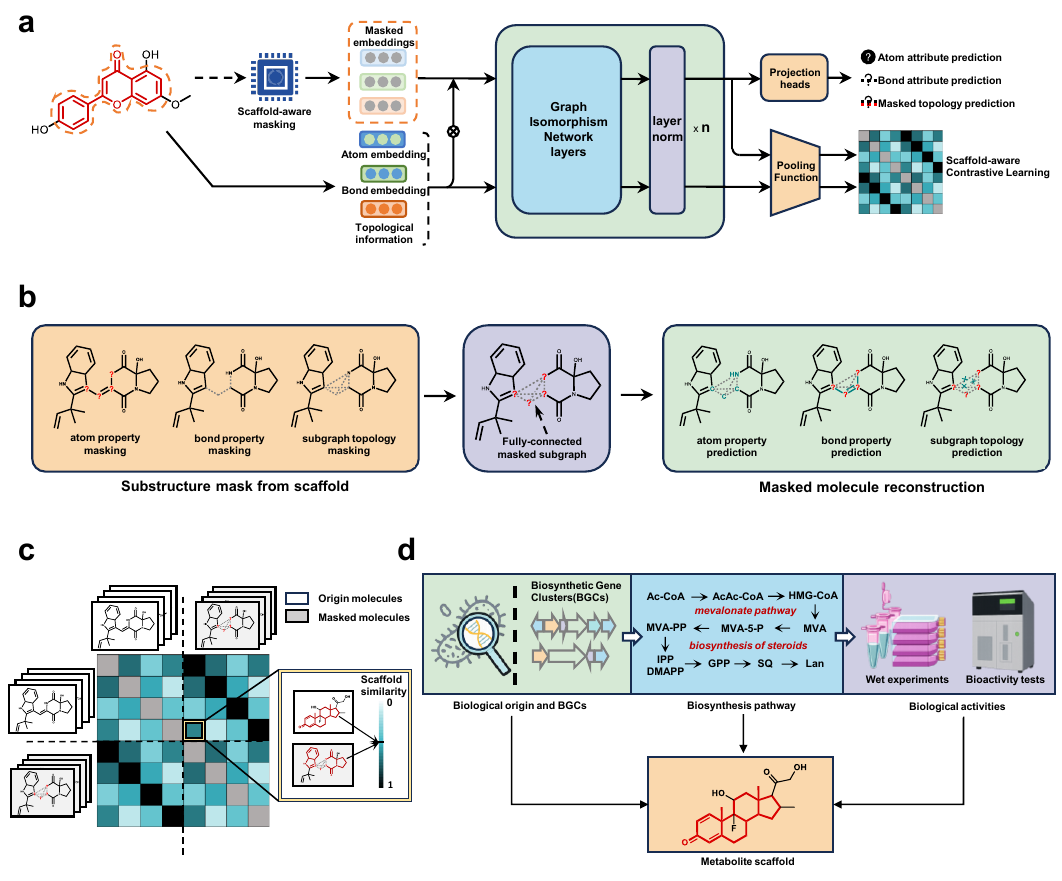}
    \caption
    {
    \textbf{Scaffold-aware pre-training framework of NaFM.}
    \textbf{(a)} Overview of NaFM pre-training. After the natural product molecule undergoes scaffold-aware masking, both the masked and unmasked information are simultaneously input into a multi-layer Graph Isomorphism Network. The masked information is then processed through a projection head to predict the masked atom attributes, bond attributes, and topological information. Meanwhile, the masked and unmasked information passes through a pooling function and is used for scaffold-aware contrastive learning. \textbf{(b)} Details of scaffold-subgraph reconstruction. First, a subgraph is randomly selected from the scaffold, consisting of multiple atoms and chemical bonds. In the subgraph, both node and edge attributes are masked, and all nodes within the subgraph are fully connected (i.e., expanded into a fully connected graph), thereby masking the topological information. During the reconstruction process, the model needs to predict both node and edge attributes, while also distinguishing between real edges and those artificially added virtual edges.   
    \textbf{(c)} Details of scaffold-aware contrastive learning. Different colors represent varying weights defined by scaffold similarity. \textbf{(d)} The "central dogma" of natural products. The biological source, biosynthetic gene clusters, biosynthetic pathways, and bioactivity of natural products are interconnected through the scaffold, which acts as a bridge linking these three key aspects. 
    } 
    \label{fig: main}
\end{figure*}

For instance, the mevalonate pathway commonly produces terpenoid and steroid scaffolds, and compounds sharing identical scaffolds often exhibit similar biochemical properties~\citep{garcia2016scaffold}. This intrinsic relationship links upstream genetic clusters and biosynthetic pathways to downstream bioactivity, as illustrated in Fig.~\ref{fig: main}(d).
Consequently, conventional molecular graph learning methods that distribute learning objectives across entire molecules are insufficient for modeling NPs. A more effective approach is to adopt a hierarchical pre-training paradigm centered on scaffolds, while capturing side-chain structures to organize the overall representation space of NPs.

Technically, current pre-training strategies face inherent limitations. In molecular contrastive learning, data augmentation typically involves masking or removing molecular substructures~\citep{wang2022molecular,liu2021pre}. However, unlike images, chemical structures are highly sensitive to such perturbations—minor structural changes can cause large variations in chemical or biological activity, a phenomenon known as the activity cliff~\citep{cruz2014activity,stumpfe2019evolving,van2022exposing}. Consequently, molecules with small structural differences cannot be treated as positive pairs~\citep{shen2023online}. Furthermore, distinct molecules may yield structurally similar or even identical negative samples, violating the assumptions of contrastive learning.
Moreover, current masked graph modeling approaches also face limitations. Masking a single atom or bond allows the model to infer missing information from adjacent nodes and edges, limiting its ability to capture global structure. Ideally, the model should consider broader molecular contexts when predicting masked components. In addition, access to complete topological information enables shortcut learning, which may weaken the intended training objectives. Studies have shown that increasing pre-training difficulty improves effectiveness~\citep{sun2022does}. Therefore, refining masking strategies and introducing additional prediction targets can substantially enrich the model’s learned representations.

In this work, we have pre-trained a foundation model for small-molecule natural products (NaFM), that seamlessly combines masked learning and contrastive learning shown in Fig.~\ref{fig: main}(a). For the masked learning task, we introduce an approach that extends beyond masking categorical features of atoms and bonds. Specifically, we also mask atomic connectivity, effectively concealing the topological structure of subgraphs. To reconstruct these subgraphs, the model must first infer bond locations and subsequently determine their types.
Recognizing that the core scaffold reflects biosynthetic logic, while structural diversity arises from enzyme-mediated modifications before or after scaffold assembly, we distinguish biosynthetic features at the scaffold level from those of peripheral substituents. Accordingly, scaffold similarity is used as a weighting function in our contrastive pre-training.
This soft-weighting strategy allows the model to distinguish strong from weak negative examples based on scaffold similarity, while incorporating informative signals from side chains.
To integrate contrastive and masked learning into a unified framework, we design the contrastive setup such that positive examples are masked molecules generated during the masked learning process, while negative examples come from other molecules and their masked variants. This mutual supervision between the two paradigms enhances representation learning and improves overall performance.

By tailoring pre-training to NPs, NaFM supports diverse downstream applications. Fine-tuned on the Natural Product Classifier (NPClassifier) dataset~\citep{kim2021npclassifier}, it achieved superior accuracy in NP taxonomy classification. NaFM embeddings effectively captured biosynthetic relationships in the NPs occurring in LOTUS database~\citep{rutz2022lotus}, clearly separating metabolites by biological origin. Using BGCs from the MiBIG repository~\citep{terlouw2023mibig}, the model enabled genomic mining directly from metabolite structures, identifying biosynthetic genes and protein families. In protein–ligand binding affinity prediction, NaFM outperformed existing models. Additionally, in a case study on acetylcholinesterase (AChE), NaFM demonstrated strong virtual screening ability, with prioritized molecules showing higher inhibitory activity than controls.

\section{Results and Discussion}\label{sec2}

\subsection{Overiew of NaFM}
NaFM is a pre-training framework tailored for natural products, integrating two tasks: scaffold–subgraph reconstruction and scaffold-aware contrastive learning.
As illustrated in Fig.~\ref{fig: main}(b), scaffold–subgraph reconstruction employs a subgraph masking strategy that extends beyond atom and bond masking to include topological information.
Because natural product scaffolds often exhibit complex polycyclic architectures with intricate atom connectivity, masking topology both enriches the learned representation and increases the reconstruction difficulty.
To further emphasize scaffold features, the subgraph mask is applied specifically to the scaffold region of each molecular graph.

In contrastive learning, transformed objects originating from the same data points are treated as positive samples, 
while those derived from different objects are treated as negative samples. 
Although this approach is effective in traditional data forms like images, its application to molecular data, particularly NPs, presents challenges. 
This complexity arises because substructures from the same molecule can exhibit significant differences after augmentation, while different molecules may share identical substructures.
To address potential mislabeling and accommodate the unique characteristics of natural products, we incorporate scaffold similarity as a soft weight, guiding the model to distinguish negative samples based on biosynthetic relationships.

To seamlessly integrate the two training tasks, the masked graphs are used not only for the scaffold-subgraph reconstruction task but also serve as positive examples in contrastive learning. In practice, contrastive learning generates a $2N\times2N$ logits matrix shown in Fig.~\ref{fig: main}(c), where the positive example for each complete molecule is its corresponding masked graph, and the negative examples are formed by other molecules and their masked graphs.
This approach fosters the learning of richer and more accurate molecular representations.

During the pre-training phase, we use approximately 0.6 million unlabeled data samples from the COCONUT database~\citep{sorokina2021coconut}. For downstream tasks, the GNN backbone parameters are retained from pre-training, while the prediction head is randomly initialized and fine-tuned together with the backbone on task-specific datasets. Further details of these processes are provided in the Methods section.

\subsection{Natural Product Taxonomy Classification}
The taxonomy classification of NPs is a classical and significant task.
A valid natural product classification system could be valuable for database mining. Predicted Taxonomy labels can be applied to infer structure-activity relationships, enabling preliminary predictions of their potential biochemical activities.


We mainly evaluate the classification performance on the NPClassifier dataset~\citep{kim2021npclassifier}, which includes seven biosynthetic pathways of NPs and detailed classifications (SuperClass and Class) based on their structures. After cleaning the dataset by removing labels and corresponding data that were too sparsely represented, we obtained approximately 77,000 entries, with labels spanning 7 types of biosynthetic pathways, 70 SuperClasses, and 563 Classes.

We first fine-tuned NaFM at the Class level. Several pre-trained molecular models were replicated and fine-tuned on the same task for comparison, with the non-pretrained Extended-Connectivity Fingerprint (ECFP) included as an additional baseline.
Datasets were split into training, validation, and test sets using stratified sampling to preserve class balance. To assess robustness under varying data scales, we further limited the number of molecules per class in the training set.
As shown in Table~\ref{ont_class}, NaFM consistently outperforms all baselines across different test sizes, with a more pronounced advantage on smaller datasets.
Graph neural network–based methods such as MolCLR, PretrainGNN, and D-MPNN show strong dependence on fine-tuning data volume, with performance dropping sharply as data become scarce. In contrast, Mole-BERT benefits from its pre-trained tokenizer, which distinguishes atoms of the same type through multiple learned tokens.
Overall, NaFM is the only pre-trained model that surpasses the rule-based ECFP fingerprint, which is the previous common choice for natural product classification.

\begin{figure*}[htbp]
    \centering
    \includegraphics[width=\textwidth]{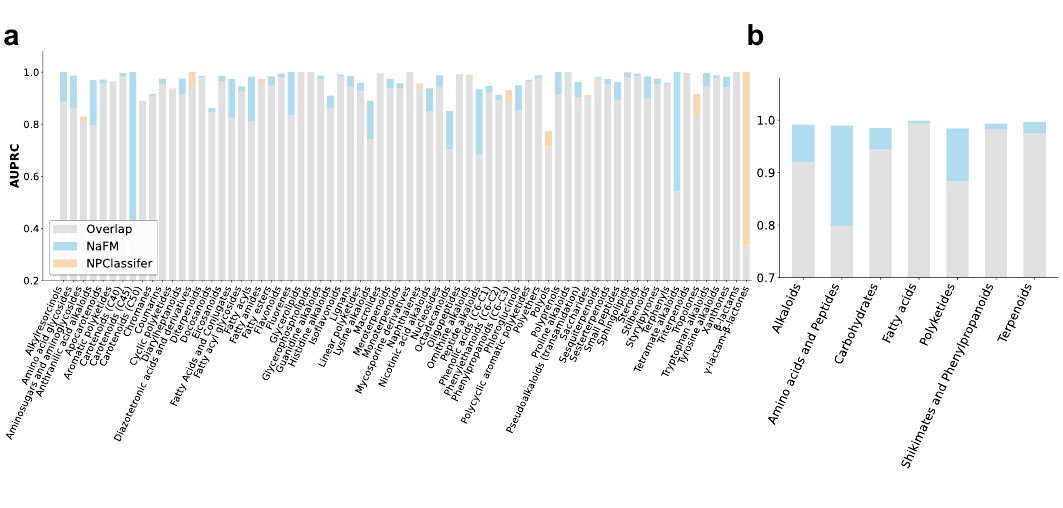}
    \caption{
    \textbf{Comparison of NaFM and NPClassifier across superclass categories and biosynthetic pathways.}
    Bar plots show model performance in terms of AUPRC for superclass categories \textbf{(a)} and biosynthetic pathways \textbf{(b)}, including carbohydrates, amino acids and peptides, alkaloids, terpenoids, shikimates and phenylpropanoids, polyketides and fatty acids. Gray segments indicate shared performance (minimum AUPRC), whereas light blue and orange segments represent the additional AUPRC achieved by NaFM and NPClassifier, respectively. 
    NaFM outperforms NPClassifier across all biosynthetic pathways and most superclass categories, with improvements of up to 24\% for amino acids and peptides. At the superclass level, NaFM achieves higher AUPRC in 53 of 71 classes, with particularly strong gains in carotenoids, alkaloids and peptide-related categories.
    }
    \label{fig: npclassifer}
\end{figure*}


Beyond comparisons with general molecular learning frameworks, we benchmarked NaFM against domain-specific tools for natural product classification. NPClassifier was selected for its open-source accessibility, scientifically grounded taxonomy, and broad adoption in the field~\citep{sorokina2021coconut,lyu2021cmnpd,van2019natural}. We fine-tuned the model on NPClassifier dataset and evaluated performance at both the Pathway and SuperClass levels. As shown in Fig.~\ref{fig: npclassifer}, NaFM consistently outperforms NPClassifier across all biosynthetic pathways and finer SuperClass categories. At the pathway level, it improves amino acid and peptide recognition by 24\%. At the SuperClass level, NaFM surpasses NPClassifier in 53 of 71 classes, achieving over 20\% gains for Carotenoids (C45), Tetramate alkaloids, Peptide alkaloids, and Anthranilic acid alkaloids. The small gap for $\gamma$-lactam-$\beta$-lactones arises from data scarcity—only 8 examples among 78,000 molecules, yielding a single test instance. These results establish NaFM as the state-of-the-art accurate framework to date for natural product classification across diverse NP databases.

\subsection{Biochemical significance of NaFM representation}

\begin{figure*}[htbp]
    \centering
    \hspace*{-0.045\textwidth}
    \includegraphics[width=1.05\textwidth]{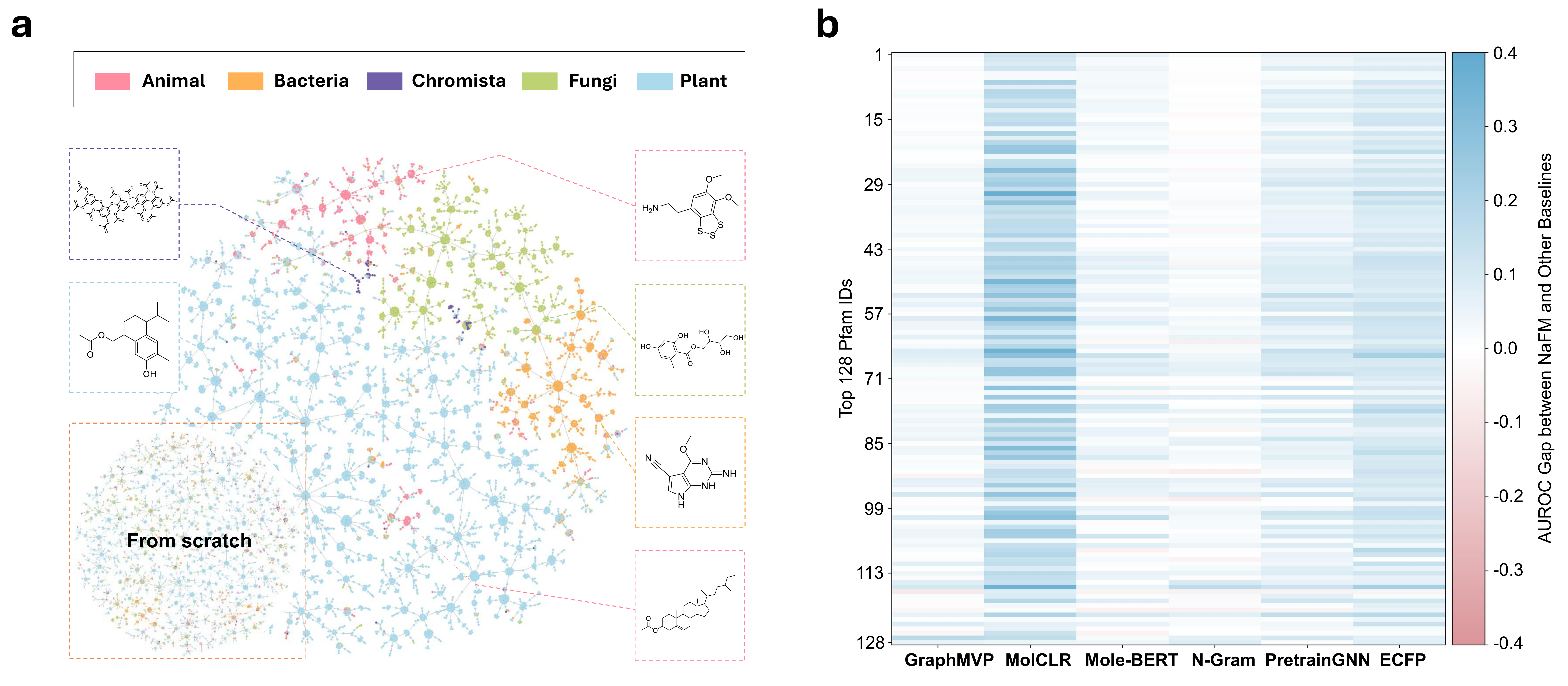}
    \caption{
        \textbf{Representation learning enables biological source discrimination and reverse BGC functional prediction.}
        \textbf{(a)} \textbf{Natural Products Atlas after representation learning.}
        Compounds from the Natural Products Atlas were embedded using the pre-trained and fine-tuned molecular representations and projected into two dimensions. The learned representation organizes chemical space into distinct clusters corresponding to biological origins (animal, bacterial, chromistan, fungal and plant), revealing clear source-specific structural signatures. Representative scaffolds are highlighted. A steroid-like scaffold in the lower right illustrates a failure case, as it occurs across multiple sources and leads to ambiguous placement. The lower-left panel shows the Atlas visualization trained from scratch without pre-training, demonstrating reduced structural discrimination.
        \textbf{(b)} \textbf{AUROC gap for predicting BGC-encoded protein families.}
        AUROC values are shown for the 128 most frequent protein families in bacteria and fungi. Color indicates the magnitude of the AUROC gap (red to blue; deeper colors represent larger absolute differences). Across most protein families, NaFM achieves higher AUROC than baseline representations, with average improvements of ~10 points over ECFP and 20–40 points over MolCLR, particularly for families with conserved functional domains such as HATPase\_c.
        }
    \label{fig: bgc_origin}
\end{figure*}

In this section, we explore the biological information integrated within our pre-trained models.
Given the strong interrelationship between molecular scaffolds, microbial species, and biosynthetic gene clusters, the specially designed pre-trained model performed exceptionally well in addressing both tasks. We fine-tune the model using the LOTUS dataset~\citep{rutz2022lotus}, which contains 130,000 data points with biological source labels.
As illustrated in Fig.~\ref{fig: bgc_origin}(a), NaFM effectively separates NPs from five major biological sources—animals, bacteria, chromista, fungi, and plants—within the representation space. 
Research on chromista is still emerging and has not kept pace with studies on other biological sources. Their unique environments and the difficulties in obtaining samples have resulted in few NPs being identified from them, with only about 1\% of data coming from chromista.
Nevertheless, despite the scarcity of training data, NaFM is still able to effectively differentiate NPs from chromista. 
For each biological source, we performed structural clustering and selected the representative scaffold corresponding to the cluster with the most data. This scaffold reflects the most characteristic structure of NPs from that source. The distinct differences in scaffolds across biological sources further highlight the strong relationship between the source and the scaffold structures.
In contrast, applying the same model architecture without pre-trained weights fails to distinguish NPs from different sources (full-sized non-pretrained Atlas shown in Supplementary Figure~S13).

In addition, we conducted a comprehensive follow-up analysis to further investigate the structural and biochemical underpinnings of the learned embeddings, as detailed in Supplementary Section~2.4.
This extended analysis includes confusion-matrix–based error inspection, identification of shared scaffolds across biological sources, and case studies of structurally ambiguous or misclassified compounds.
These results reveal that most misclassifications stem from genuine biochemical overlap—such as common terpenoid and indole scaffolds shared by plants, fungi, and animals. In addition, we generated atlases using different baselines under various training scenarios to further validate NaFM’s ability to classify biological sources (see Supplementary Figure~S16). Moreover, to demonstrate NaFM’s capability in distinguishing finer-grained biological classifications, we also produced a heatmap combined with a phylogenetic tree organized at the levels of domain, kingdom, phylum, and class (see Supplementary Figure~S17). These results provide strong evidence that NaFM effectively captures biologically meaningful distinctions.

Genetic studies have shown that bioactive natural products are synthesized by co-localized genes organized into biosynthetic gene clusters (BGCs)~\citep{martin1989organization,martin1992clusters}. While traditional methods predict metabolites from BGC structures~\citep{hannigan2019deep,carroll2021accurate,sanchez2023expansion}, reverse prediction—from metabolite structures to BGCs—enables precise localization of genomic regions responsible for biosynthesis, improving cluster selection and large-scale production.
To evaluate this capability, we assembled a dataset of over 2,000 bacterial and fungal BGCs from MIBiG~\citep{terlouw2023mibig} and their associated protein families from Pfam~\citep{mistry2021pfam}. NaFM and several baseline molecular representations were trained to predict BGC-encoded protein families from metabolite structures. As shown in Fig.~\ref{fig: bgc_origin}(b), we analyzed 128 frequent protein families (Supplementary Table S10); NaFM outperforms MolCLR, ECFP, and PretrainGNN in nearly all categories.

Specifically, analyzing the blue blocks in the figure, the model shows significantly better classification performance on certain protein families, such as the Conserved Protein Domain Family (ID: 65 in the figure) and HATPase\textunderscore{}c (ID: 115 in the figure), compared to other baselines. The Conserved Protein Domain Family includes protein domains that are widely conserved across different species and are involved in a variety of biological processes, indicating structural and functional similarity~\citep{marchler2007cdd}, while the HATPase\textunderscore{}c family is associated with ATP hydrolysis, playing a key role in cellular processes like signal transduction and DNA repair~\citep{ulrich2010mist2}. Both of these families are considered conserved protein families. These results suggest that NaFM is capable of making effective predictions by learning the conserved features within these structures. In addition to analyzing the top 128 protein families, we also compared the average classification performance across all protein families. As shown in Supplementary Figure~S15, NaFM significantly outperforms the other baselines in terms of both AUROC and AUPRC, with a smaller variance. 

Furthermore, to demonstrate the applicability of NaFM’s genome-mining capabilities in real-world scenarios, we conducted a comprehensive case study in which NaFM was applied to the reverse inference of biosynthetic features of natural products. Detailed experimental procedures and corresponding results are provided in Supplementary Section 2.5.

In summary, NaFM demonstrates exceptional genetic-level predictive performance, establishing it as a powerful tool for future BGC mining and biosynthetic pathway analysis, particularly for natural products that are difficult to synthesize or annotate.

\subsection{Natural product bioactivity prediction and virtual screening}

\begin{figure*}[htbp]
     \centering
     \hspace*{-0.1cm}
     \includegraphics[width=0.97\textwidth]{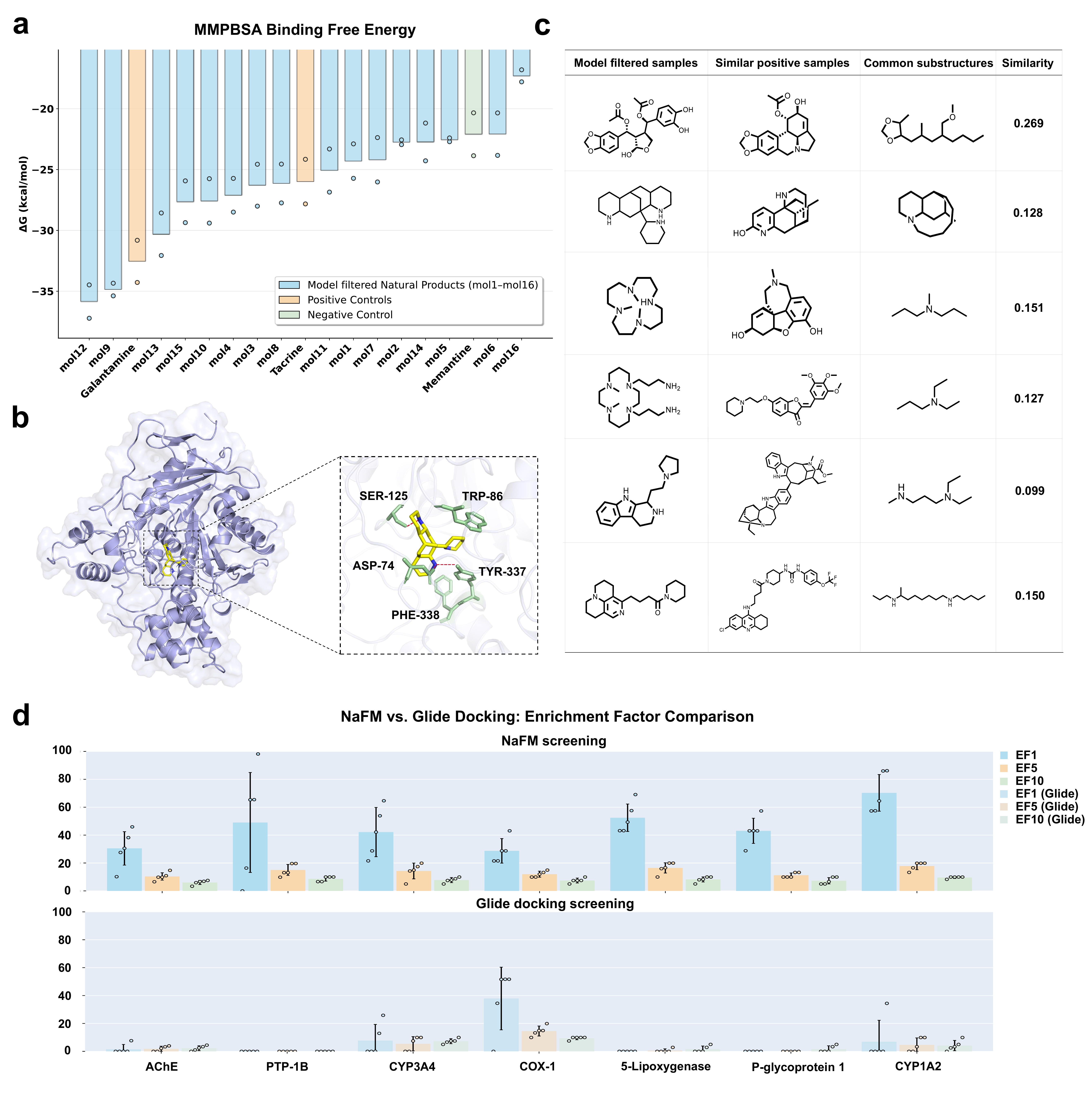}
     \caption
     {
     \textbf{Computational validation and benchmarking of NaFM-predicted inhibitors.}
     \textbf{(a)}Binding free energies ($\Delta G$) of top 16 predicted compounds (blue), positive controls (orange: galantamine, tacrine), and negative control (light green: memantine). $\Delta G$ was estimated via MMPBSA (GB model) using 100 snapshots from the final 10 ns of equilibrated trajectories. Bar heights represent the mean of $n=2$ independent MD simulations (individual data points overlaid), performed as methodological repeats to ensure stability.
     \textbf{(b)} Visualization of MD-simulated interactions. In the ligand, carbon atoms are shown in bright yellow and nitrogen atoms in blue. Surrounding amino acid residues are colored green. The model-selected molecule forms three hydrogen bonds with TYR337, ASP74, and SER125 (red dashed lines), and exhibits additional close-contact interactions with other residues such as TRP86 and PHE338.
     \textbf{(c)} The first and second columns display several high-scoring molecules predicted by the model, along with their most similar experimentally validated actives encountered during the fine-tuning process. The third and fourth columns illustrate the maximum common substructures (MCS) between each pair of molecules and their corresponding Tanimoto similarities computed using ECFP4 fingerprints.
     \textbf{(d)} Enrichment factors for NaFM and Glide across seven targets. Results represent mean ± s.d. from five-fold cross-validation ($n=5$ independent splits). Bar heights and error bars denote mean and standard deviation, respectively. Higher values indicate superior screening performance.
     }
     \label{fig: virtual_screening}
\end{figure*}

Due to the difficulty in synthesizing or extracting natural products, there is significantly less activity data available for them compared to synthetic molecules~\citep{zeng2024natural}. Consequently, the knowledge learned through fine-tuning is limited, posing a challenge for the amount of information the model can acquire directly during pre-training. Although cell line data are more abundant, their complexity—stemming from multiple proteins, organelles, and signaling pathways—renders the structure-cytotoxicity relationship intricate. 
In contrast, protein targets with well-characterized 3D structures and annotated binding sites enable clearer structure–activity relationships through defined molecular interactions (e.g., hydrogen bonds, hydrophobic contacts, and ionic bonds). Therefore, we focused on such structurally resolved protein targets for subsequent activity prediction and screening experiments.

To validate the activity prediction capability of NaFM, we selected human protein targets and HIV type-1 reverse transcriptase activity data from the NPASS natural product activity database, with activity data greater than 100 for each target. We then fine-tuned NaFM and various molecular representation baselines on this refined dataset and recorded the Root Mean Square Error(RMSE) for each method across different protein target activity prediction tasks. The results are shown in Table~\ref{regression_rmse}. As demonstrated, NaFM surpasses all baselines, with the exception of N-Gram on HIV type-1 reverse transcriptase, where NaFM performs marginally less effectively. 
The slightly weaker performance of NaFM on HIV-1 reverse transcriptase likely originates from the distinct characteristics of the molecular features. The N-Gram baseline, which encodes short, local SMILES substrings and substructure frequency patterns, effectively exploits recurring motif-like signals characteristic of certain antiviral chemotypes. In contrast, NaFM’s ligand-based self-supervised embeddings primarily capture higher-order scaffold topology and broader functional-group context, accounting for its state-of-the-art performance across most human protein targets. Importantly, even for HIV-1 reverse transcriptase our model still outperforms most baselines and remains competitive.

Moreover, NaFM substantially outperforms GraphMVP and PretrainGNN across diverse protein targets, including PTP1B, AChE, COX-2, Tyrosinase, and CYP3A4. NaFM consistently yields lower RMSE values, such as 0.9286 on COX-1 compared with 3.0083 for GraphMVP and 2.6831 for PretrainGNN, with the advantage most pronounced for complex or data-scarce targets.

The weaker performance of GraphMVP likely stems from its reliance on molecular geometry during pre-training. As discussed in Supplementary Section~2.2, natural products combine rigid substructures with flexible global conformations, resulting in broad and diverse 3D distributions. This structural heterogeneity hampers geometry-based models like GraphMVP, limiting their ability to capture the full conformational space of natural products.
We further applied NaFM’s representations to virtual screening of natural products—a critical yet challenging step in drug discovery~\citep{maia2020structure}. Conventional docking and deep learning methods~\citep{friesner2004glide,trott2010autodock,kimber2021deep} are effective for synthetic compounds but often struggle with NPs, whose complex stereochemistry and reactive groups complicate synthesis and screening~\citep{ma2011molecular}. 

To evaluate NaFM’s robustness across functionally and structurally diverse targets, we performed a systematic validation on acetylcholinesterase (AChE), a key esterase in neurodegenerative disease therapy~\citep{soreq2001acetylcholinesterase}. In addition, we carried out a supplementary validation on protein tyrosine phosphatase 1B (PTP-1B)~\citep{delibegovic2024protein}, in Supplementary Section~2.6.

For the AChE case study, we utilized the IC$_{50}$ data from the NPASS database, categorizing compounds as active or inactive based on a micromolar cutoff. 
The model was fine-tuned on the activity data and subsequently used to score the activity of compounds in the COCONUT dataset. Following the screening strategy above, we performed 100 ns MD simulations and binding free energy ($\Delta G$) calculations for the top 16 model-ranked compounds and three reference drugs: the AChE inhibitors galantamine and tacrine, and memantine. Memantine, which is co-administered clinically but does not bind or inhibit AChE, was included as a negative control.
Docking poses were used as initial conformations, and 100 ns MD simulations were performed with GROMACS using the AMBER99SB-ILDN force field. Representative RMSD profiles are shown in Supplementary Fig. S14. Binding free energies ($\Delta G$) were estimated by MMPBSA on the equilibrated trajectory segments (final 10 ns after RMSD stabilization; Fig. \ref{fig: virtual_screening}(a)). Two of the 16 compounds exhibited lower $\Delta G$ than both positive controls, indicating stronger predicted binding, and the consistently low $\Delta G$ values across top-ranked molecules confirm NaFM’s screening effectiveness. Comparable PTP-1B results are reported in Supplementary Section 2.6.

For the top-ranked compound Mol12 (lowest $\Delta G$), an equilibrated MD snapshot was analyzed in PyMOL to visualize binding (Fig. \ref{fig: virtual_screening}(b)). Mol12 forms a dense interaction network in the AChE active-site gorge: its ammonium group (NH$_2^+$) engages ASP74, SER125, and TYR337 via hydrogen bonds and is further stabilized by hydrophobic and van der Waals contacts with TRP86 and PHE338.
Beyond potency, we evaluated scaffold novelty. Structural similarity to known actives was quantified using MCS overlap and ECFP4 Tanimoto similarity (Fig. \ref{fig: virtual_screening}(c)). Despite a diverse fine-tuning set (44 actives, 37 Bemis–Murcko scaffolds), NaFM recovered active compounds with low similarity to known ligands, demonstrating strong generalization and the ability to discover novel scaffolds.

To further demonstrate NaFM’s capability across diverse human protein targets, we fine-tuned the model on activity data from seven representative targets and applied the same virtual screening procedure, using enrichment factors (EFs) as metrics. Glide, a widely used docking-based screening tool, was used for comparison. As shown in Fig.~\ref{fig: virtual_screening}(d), NaFM outperforms Glide in EF1\%, EF5\%, and EF10\% across all targets except COX-1, where EF1\% is slightly lower. Notably, for AChE, PTP-1B, 5-lipoxygenase, and P-glycoprotein 1, Glide shows minimal enrichment, whereas NaFM achieves excellent performance, likely reflecting the difficulty of fitting complex NP structures and scaffold geometries.

These results provide multi-target, physics-informed validation of NaFM’s transferability across protein families. While broader cross-family benchmarking remains future work, NaFM demonstrates strong potential as an efficient virtual screening framework for discovering bioactive and structurally diverse natural products.

\subsection{Ablation and Sensitivity Analysis.}

We conducted a comprehensive ablation analysis to quantify the contributions of different components in NaFM, including its architectural modules, contrastive learning design, and dataset composition.

\paragraph{Model Architecture}
To assess the individual effects of masked graph modeling and contrastive learning, we retrained NaFM while removing each module independently (Table S3). The complete NaFM achieved the highest performance (AUPRC 70.10 ± 0.92; accuracy 64.78 ± 0.80). Removing masked graph modeling reduced accuracy to 57.87 ± 2.13 vs. 64.78 ± 0.80 (\( -6.9 \) points), while excluding contrastive learning led to 61.75 ± 2.05 vs. 64.78 ± 0.80 (\( -3.0 \) points). These results confirm the complementary nature of the two objectives and highlight contrastive learning as the most critical component. We also examined the influence of the encoder on model performance. Detailed results are provided in Supplementary Tables S4–S5.


\paragraph{Contrastive Learning Design.}
To further investigate the contribution of NaFM’s contrastive learning design, we performed ablation experiments focusing on two key components: the incorporation of scaffold-based weighting and the choice of scheduling strategies (Table S5). Compared with a standard non-weighted contrastive setup (AUPRC 65.24 ± 0.86 at 4 samples/class), incorporating scaffold-based weighting and dynamic scheduling improved AUPRC by nearly 5 points. Moreover, alternative scheduling strategies such as exponential (65.64 ± 0.43) or logarithmic (67.12 ± 0.36) vs. cosine (70.10 ± 0.92) consistently underperformed. These findings emphasize that the synergy between scaffold-aware weighting and cosine-based scheduling is key to NaFM’s robust contrastive representation learning.

\paragraph{Pre-training Dataset Composition.}
To disentangle the effects of data from methodological gains, we re-pretrained all baseline methods that support pre-training (MolCLR, PretrainGNN, Mole-BERT) on the same COCONUT dataset and evaluated them across varying data-density regimes for classification tasks (Table S6). Re-pretrained MolCLR underperformed its original checkpoint in both low- and high-data regimes (41.49 ± 1.05 vs. 45.76 ± 1.98), while PretrainGNN also showed decreased performance at high data densities (81.28 ± 0.83 vs. 87.82 ± 0.13). Mole-BERT exhibited an increase in high-data performance (89.47 ± 0.67 vs. 85.69 ± 0.70), but its low-data performance dropped noticeably (61.07 ± 0.58 vs. 66.32 ± 1.23). Notably, regardless of whether these models were pre-trained on COCONUT, all baselines performed substantially worse than NaFM across all data regimes. These findings indicate that the observed advantages do not arise from dataset choice or data bias. Instead, they underscore that the key driver of NaFM’s effectiveness lies in the suitability of its pre-training objectives to the natural product domain.

Overall, these ablations demonstrate that NaFM’s superior performance stems from the combined effects of (i) its dual pre-training objectives, (ii) scaffold-aware and dynamically scheduled contrastive learning, and (iii) a pre-training strategy purpose-built for natural product chemistry.

\section{Conclusion}\label{sec3}
In this paper, we have proposed NaFM, a foundation model for small molecule natural products. 
NaFM leverages the power of a large amount of unlabeled natural product structures and an efficient biological insight-inspired pre-training strategy.
Our method has been evaluated across a range of tasks, including upstream genome mining, biological source identification, structural classification, metabolic pathway prediction, and downstream activity prediction and screening.
On most of the benchmarks, NaFM outperforms both traditional molecular fingerprints and deep-learning based pre-training frameworks.
In the future, NaFM holds great potential as a foundation model for natural product based drug discovery and biosynthesis studies. Furthermore, it can serve as a tool-driven platform that streamlines key aspects of research workflow for natural products, integrating computational analysis, predictive modeling, and experimental validation to accelerate discoveries in the field.

\section{Methods}\label{sec4}

\subsection{Initial Representations of molecules}

In our work, a molecule is represented as an attributed graph $G=(V,E)$, where atoms are nodes ($V$) and chemical bonds are edges ($E$). We adopt the standard feature representation commonly used in previous studies~\citep{wang2022molecular}. As detailed in Supplementary Table~S1, the initial node features are constructed by summing the embeddings of attributes including atom types, chirality, and formal charges. Similarly, the initial bond features are obtained by summing the embeddings of bond types and bond directions.
\begin{equation}
    \text{Representation}_{atom/bond}=\sum_{i=0}^{F_{atom/bond}}\text{Embed}_i\left({\text{Attribute}_i}\right)
\end{equation}

where, $F_{*}$ is the number of atom or bond features. These initial features also serve as labels for the reconstruction task. We define the possible categories as the Cartesian product of all attribute sets, resulting in $9 \times 4 \times 5 = 180$ distinct atom categories ($M_{atom}$) and $ 4 \times 3 = 12$ distinct bond categories ($M_{bond}$). In other words, each unique combination of atom attributes forms an atom category, and each unique pair of bond attributes defines a bond category.

\subsection{Model Architecture}

We adopt the traditional message-passing scheme introduced in GIN~\citep{xu2018powerful}, with modifications to incorporate edge features as proposed in \citep{hu2021ogb}. The node representation is updated as follows:

\begin{equation}
    \hat{h}^{l+1}_i = \text{MLP}\left(h^{l}_i + \sum_{j \in \mathcal{N}(i)} (h^{l}_j + e_{ij})\right),
\end{equation}

where \(h^{l}_i\) is the representation of node \(i\) in the \(l\)-th layer, \(e_{ij}\) represents the edge feature between nodes \(i\) and \(j\), and \(\mathcal{N}(i)\) denotes the neighbors of node \(i\). To stabilize training, we apply layer normalization after each GIN block:

\begin{equation}
    h^{l+1}_i = \text{LayerNorm}(\hat{h}^{l+1}_i)
\end{equation}

The final node representations, \(h^{L}\), are aggregated using an average pooling operation to derive the graph-level representation.
\begin{equation}
    h_{\text{graph}} = \frac{1}{|V(G)|}\sum_{v\in G}h_v
\end{equation}
where \(|V(G)|\) is the number of nodes in each graph.
For the contrastive learning task, an additional MLP is used to project the graph representation into the latent space. For reconstruction, the final node representations are mapped to the corresponding atom categories, while the final edge representations are obtained by concatenating the node representations and applying an MLP to adapt them to the specific tasks.
During fine-tuning on downstream tasks, we remove the contrastive learning and reconstruction head and attach an MLP to the final graph representation for classification or regression tasks.

\subsubsection{Scaffold-aware Contrastive Learning}

The issue of mislabeling in contrastive learning can be mitigated through the application of re-weighting techniques~\citep{wang2022improving}. However, determining appropriate weights is a challenging task. Interestingly, the distinctive characteristics of natural products—specifically, the strong correlation between their scaffold structures and their properties—offer a promising solution for weight selection: scaffold similarity.
To implement this, we compute the cosine similarity between the MACCS fingerprints~\citep{durant2002reoptimization} of their scaffolds. Specifically, scaffolds are defined here according to the Bemis–Murcko framework. This similarity is subsequently incorporated into the contrastive learning loss function for re-weighting, as detailed below:
\begin{equation}
    \mathcal{L}_{c}^{w}=\frac{1}{N}\sum_{m=1}^{N}\left[\ell\left(2m-1,2m\right)+\ell\left(2m,2m-1\right)\right], 
    \ell_{i,j} = -\log\frac{s_{i,j}}{s_{i,j} + \sum_{k=1}^{2N}\mathbb{I}_{[k \neq i,j]}(1-w_{i,k})s_{i,k}}
\end{equation}
where $s_{i,j}=\exp\left(\text{sim}(h_i, h_j)/\tau\right)$ is the similarity in exponential space between the representations of $i$-th molecule and $j$-th molecule in a batch contained $N$ samples and $\mathbb{I}$ is the indicator function.

The weighted loss function can be interpreted as assigning lower weights to negative sample pairs from different molecules that exhibit more scaffold similarity. This allows the model to reduce its attention to these "false negatives", preventing it from being misled.
As a result, the model achieves a more robust representation of molecular topology structures. 
Although reweighting techniques were introduced, the model tended to overfit the contrastive learning task as training progressed, leading to attempts to distinguish these "false negatives". To further address this issue, we developed a cosine loss weight scheduler as follows:
\begin{equation}
\lambda_{c}^{i}=\frac{1}{2}\cos(1 + \frac{i\pi}{N})
\end{equation}
where $i$ is the current epoch and $N$ is the total epochs. In this way, the model can mitigate overfitting on this task while ensuring sufficient training for other tasks.

\subsubsection{Scaffold-subgraph Reconstruction}

With the data-augmented samples, which naturally provide labels for mask modeling, the two frameworks were seamlessly integrated. These two approaches are complementary: contrastive learning is prone to overfitting, as contrastive models may find shortcuts through trivial representations. However, with the aid of mask modeling, the model goes beyond merely distinguishing between positive and negative sample pairs. It leverages information from functional groups (non-masked) to predict scaffold information (local). Simultaneously, contrastive learning enhances the model by helping it learn a global semantic representation.
It is important to note that we define the task as scaffold-subgraph reconstruction, while discarding traditional approaches such as atomic masking and bond deletion. 
These tasks are considered too simplistic for the model and do not contribute meaningful knowledge. 
To ensure that the model focuses on scaffold modeling, we limit the masked portions exclusively to the atoms within the scaffold. 
In the reconstruction phase, we introduce a loss function that combines three classification tasks: atom type prediction, link prediction, and bond type prediction.
\begin{align}
    \mathcal{L}_{r} &= \mathcal{L}_{atom} + \mathcal{L}_{link} + \mathcal{L}_{bond} \\ &= -\frac{1}{N_{atom}}\sum_{i}\sum^{M_{atom}}_{c=1}y_{ic}\log(p_{atom}^{ic}) - \frac{1}{N_{bond}}\sum_{j}\sum^{2}_{c=1}y_{jc}\log(p_{link}^{jc}) -\frac{1}{N_{bond}}\sum_{j}\sum^{M_{bond}}_{c=1}y_{jc}\log(p_{bond}^{jc})
\end{align}
where, $N_{*}$ is the number of atoms or bonds, $M_{*}$ is the number of atom or bond category.
Here, $y_{ic}$ represents the indicator function, where the value is 1 if sample $i$ belongs to category $c$, and 0 otherwise. Additionally, $p^{ic}_{*}$ denotes the probability that the model predicts sample $i$ belongs to category $c$. More detailed information about atom or bond categories is shown in Methods section.

\subsection{Scaffold Masking Algorithm}

Scaffold masking is a critical step in our framework, as it plays a key role in both the augmented data of contrastive learning and the reconstruction task.
The masking process is described in detail below.

As shown in Algorithm~\ref{alg: subgraph}, initially, one scaffold atom is randomly selected, and its corresponding node is removed. 
The process continues by identifying the neighbors of the currently selected nodes, prioritizing those in the scaffold. 
These neighboring nodes are shuffled and iteratively removed until the desired number of nodes is reached or no further eligible neighbors exist. 
The output is a tensor containing the indices of the removed nodes. After that, all removed nodes are marked as masked, and all edges containing any of the removed nodes are also masked to form a augmented graph of contrastive learning and input/output pair of the reconstruction task.

\vspace{0.5cm}
\begin{algorithm}[H]
\caption{Masked Subgraph Generation}
\label{alg: subgraph}
\KwIn{Graph $G$, Scaffold IDs $S$, Mask ratio $m$}
\KwOut{Modified graph $G_{\text{new}}$, Masked nodes $V_{\text{masked}}$}

\textbf{Initialization:} \\
\Indp
$ M \leftarrow \max(\min(\lceil |V(G)| \cdot m \rceil, |S|), 2)$ \\
$V_{\text{masked}} \leftarrow \emptyset$ \\
$V_{\text{temp}} \leftarrow \text{Random choice}(S, 1)$ \\
\Indm

\While{$|V_\text{masked}| < M$}{
    $V_\text{neighbors} \leftarrow \emptyset$ \\
    \ForEach{$n \in V_\text{temp}$}{
        $V_\text{neighbors} \leftarrow V_\text{neighbors} \cup \{i \in \text{Neighbors}(G, n) \mid i \notin V_\text{temp} \text{ and } i \in S\}$ \\
    }
    $V_\text{neighbors} \leftarrow \text{Shuffle}(V_\text{neighbors})$ \\
    \ForEach{$n \in V_\text{temp}$}{
        \If{$|V_\text{masked}| < M$}{
            $G \leftarrow \text{Remove node}(G, n)$ \\
            $V_\text{masked}\leftarrow V_\text{masked} \cup \{n\}$ \\
        }
        \Else{
            \textbf{break}
        }
    }
    $V_\text{temp} \leftarrow \text{unique nodes in } V_\text{neighbors}$ \\
    \If{$V_\text{temp} = \emptyset$}{
        \textbf{break}
    }
}
\Return{$V_{\text{masked}}$}
\end{algorithm}


\subsection{Training Details}

\paragraph{Pre-training}
We implement a 6-layer Graph Isomorphism Network (GIN)~\citep{xu2018powerful} with layer normalization as the GNN backbone, using a hidden dimensionality of 512. An average pooling operation is applied as the readout function to extract a global molecular representation of 1024 dimensions. For the contrastive learning (CL) task, a single MLP with LeakyReLU activations projects the representation to 256 dimensions. For the reconstruction task, three MLPs with LeakyReLU activations map the node or edge representations to their respective output dimensions. All MLPs use a hidden dimensionality of 512. The masking ratio is set to 0.2. 

We optimize the weighted CL and reconstruction loss using the AdamW optimizer with a weight decay of \(10^{-5}\). The temperature for the CL loss $\tau$ is set to 0.1. A cosine decay schedule is applied to the scaling factor of the CL loss, decreasing from 1 to 0. The learning rate is set to \(1 \times 10^{-4}\) with a cosine decay schedule. The model is trained with a batch size of 256 for a total of 300 epochs. To avoid overfitting, we also set the dropout ratio to 0.1. A detailed hyper-parameter search space is provided in Supplementary Table~S2. To pre-train NaFM, we employed four NVIDIA A100 GPUs and trained the model continuously for five days.

\paragraph{Downstream Fine-tuning}

For downstream task fine-tuning, we replace the original contrastive learning (CL) head and reconstruction head with an additional, randomly initialized MLP on top of the GNN backbone.
The softmax cross-entropy loss is used for classification tasks. For each task, we use the AdamW optimizer with a weight decay of \(1 \times 10^{-5}\) and a learning rate of \(5 \times 10^{-4}\), along with a cosine learning rate decay schedule. The model is fine-tuned for 300 epochs with a batch size of 512. 
The dropout ratio in the added MLP is set to 0.3. 
Early stopping is applied if the evaluation metric does not improve for 20 consecutive epochs. 
Thanks to the robust performance of the pre-training, we do not perform hyper-parameter tuning for the downstream tasks.
Regarding the data splitting strategy, we employed stratified splits for the taxonomy classification of natural products to ensure balanced class distribution across the training, validation, and test sets. For the biological source, Pfam prediction, and bioactivity prediction tasks, we used random splits.
For all pre-trained baselines, we use the same settings as described above.
For baselines trained from scratch, we follow their default configurations.

\section{Data availability}
The minimum datasets required to interpret, verify, and extend the findings of this study are publicly available at Figshare \url{https://doi.org/10.6084/m9.figshare.28980254.v1}(ref.\citep{Ding2025_figshare}). The repository contains the following datasets used for pre-training and downstream evaluation:

\begin{itemize}
    \item \textbf{pretrain\_smiles.pkl}: Preprocessed data used for model pre-training. The original data was obtained from the COCONUT database: \url{https://coconut.naturalproducts.net/}
    \item \textbf{classification\_data.csv}: Data prepared for the Natural Product Taxonomy Classification experiment. The original dataset was sourced from: \url{https://doi.org/10.26434/chemrxiv.12885494.v1}(ref.\citep{NPClassifier2020})
    \item \textbf{NPClassifier\_dataset\_refreshed.csv}: Data curated for direct comparison with NPClassifier. Original data is available at: \url{https://github.com/mwang87/NP-Classifier/tree/master/training/Data/NPClassifier_dataset.xlsx}
    \item \textbf{regression\_data.csv}: Dataset used for natural product bioactivity prediction tasks. The original data was retrieved from the NPASS database: \url{https://bidd.group/NPASS/}
    \item \textbf{lotus\_data.csv}: Data prepared for biological source prediction and mining. The source data was collected from the LOTUS database: \url{https://lotus.naturalproducts.net/}
    \item \textbf{bgc\_data.csv}: Dataset constructed for biosynthetic gene cluster mining. The original sources include the MIBiG database (\url{https://mibig.secondarymetabolites.org/}) and Pfam (\url{http://pfam.xfam.org/})
    \item \textbf{external\_data.csv}: Dataset used for bioactivity screening of natural products. The original data was obtained from the NPASS database: \url{https://bidd.group/NPASS/}
\end{itemize}
Also, the pre-trained weights are provided at Zenodo \url{https://doi.org/10.5281/zenodo.15385335}(ref.\citep{Ding2025_NaFM_Weights})

\section{Code availability}
The source code used in this study is publicly available at \url{https://github.com/TomAIDD/NaFM-Official}.
An archived version corresponding to the code used in this manuscript has been deposited in Zenodo (ref.\citep{Ding2025_NaFM_Code})

\section{Acknowledgements}
We thank all members of the collaborative team for their contributions to this work. This research was supported by the National Key Research and Development Program of China (Grant Nos. 2022YFC2804900 to Aili Fan), the Fundamental and Interdisciplinary Disciplines Breakthrough Plan of the Ministry of Education of China (Grant Nos. JYB2025XDXM504 to Nanning Zheng), the National Natural Science Foundation of China (Grant Nos. 22277006 to Zhenming Liu, 92259302 to Zhenming Liu), and the Beijing Natural Science Foundation (Grant Nos. 7242195 to Zhenming Liu).

\section{Author contributions}
Y.D. and B.Q. conceived the project. Y.D. and Y.W. processed the dataset and trained the model, with both performing computational downstream tasks and Y.D. also conducting biochemical analyses. Y.Z. and J.Y. provided support for specific downstream datasets. Y.D. and B.Q. analyzed the results, and Y.D. wrote the original draft of the manuscript. Y.D., B.Q., S.L., J.Y., C.S., Q.L., and L.Z. contributed to manuscript revisions. The project was supervised by Z.L., N.Z., Y.W., and B.Q. All authors participated in discussions and provided critical feedback.

\section{Competing interests}
The authors declare no competing interests.

\clearpage
\section{Tables}

\begin{table*}[htbp]
    \caption
    {
    \textbf{The AUPRC results for natural product taxonomy classification compared to other baselines.}
    AUPRC values were computed three times using different random seeds.
    }
    \begin{threeparttable}
    \label{ont_class}
    \centering
    \resizebox{\linewidth}{!}
    {
    \begin{tabular}{@{\extracolsep{\fill}}ccccccc@{\extracolsep{\fill}}}
        \toprule
        \makecell{Training samples \\per Class} & 4  & 8  & 16  & 24  & 40  & 64 \\
        \midrule
        N-Gram~\citep{liu2019n} & $44.72\pm1.91$ & $56.61\pm0.66$ & $66.36\pm1.57$ & $71.11\pm0.77$ & $73.34\pm1.14$ & $75.77\pm0.54$\\
        PretrainGNN~\citep{hu2019strategies} & $44.83\pm0.82$ & $61.85\pm0.78$ & $75.76\pm0.62$ & $80.50\pm0.17$ & $85.31\pm0.36$ & $87.82\pm0.13$ \\
        D-MPNN~\citep{yang2019analyzing} & $46.63\pm0.23$ & $60.88\pm0.48$ & $75.73\pm0.12$ & $80.96\pm0.24$ & $86.64\pm0.44$ & $89.23\pm0.79$\\
        MolCLR~\citep{wang2022molecular} & $45.76\pm1.98$ & $65.80\pm1.51$ & $78.14\pm1.08$ & $83.20\pm0.74$ & $85.56\pm0.11$ & $88.22\pm0.34$ \\
        Mole-BERT~\citep{xia2023mole} & $
        66.32\pm1.23$ & $73.39\pm0.66$ & $
        78.25\pm0.45$ & $80.83\pm0.51$ & $
        83.57\pm0.59$ & $85.69\pm0.70$ \\
        ECFP~\citep{rogers2010extended} & $69.17\pm0.19$ & $78.21\pm0.79$ & $83.82\pm0.37$ & $86.28\pm0.49$ & $88.52\pm0.48$ & $89.75\pm0.45$\\
        GraphMVP~\citep{liu2021pre} & $64.50\pm0.74$ & $78.41\pm0.09$ & $85.71\pm0.43$ & $87.88\pm0.26$ & $89.72\pm0.45$ & $91.07\pm0.43$ \\
        NaFM& $\mathbf{70.10\pm0.92}$  & $\mathbf{79.89\pm0.07}$ & $\mathbf{87.37\pm1.51}$ & $\mathbf{89.15\pm0.22}$ & $\mathbf{90.77\pm0.26}$ & 
        $\mathbf{91.75\pm0.47}$\\
        \bottomrule
    \end{tabular}
    }
    \end{threeparttable}
    \begin{flushleft}
    \small \textit{Note}: Results are reported as mean ± standard deviation. Bold values indicate the best performance in each column.
    \end{flushleft}
\end{table*}

\begin{table*}[htbp]
    \caption
    {
    \textbf{Comparison of RMSE performance for natural product bioactivity regression.}
    Bioactivity values are expressed as $pIC_{50}$. Reported RMSE values represent the mean of three independent runs initialized with different random seeds.}
    \begin{threeparttable}
    \label{regression_rmse}
    \centering
    \resizebox{\linewidth}{!}
    {
    \begin{tabular}{@{\extracolsep{\fill}}lcccccccc@{\extracolsep{\fill}}}
        \toprule
        Target Name &	PTP1B &	AChE &	COX-2 &	HIV type-1 RT & Tyrosinase	& CYP3A4 & 	MRP4	& COX-1 \\
        Number of samples &	612	&341&	190&	186	&186	&178	&177	&140 \\
        \midrule
        N-Gram~\citep{liu2019n} & $0.8493\pm0.2623$	& $1.3774\pm0.3995$	& $1.0022\pm0.3192$	& $\mathbf{1.0606\pm0.1613}$	& $0.7822\pm0.1408$ &	$0.7846\pm0.2037$ &	$0.2747\pm0.0846$ &	$0.9529\pm0.1643$ \\
        PretrainGNN~\citep{hu2019strategies} & $0.8652\pm0.2110$ &	$1.5462\pm0.2265$ & 	$1.7231\pm0.3754$ &	$1.5210\pm0.2606$ & 	$1.6957\pm0.2915$ & 	$1.4157\pm0.2923$ & 	$1.6869\pm0.2476$ &	$2.6831\pm0.2056$ \\
        D-MPNN~\citep{yang2019analyzing} & $0.8577\pm0.2165$ &	$1.2881\pm0.1992$ &	$1.0272\pm0.2468$ &	$1.1826\pm0.1683$ &	$0.8206\pm0.1891$ &	$0.6946\pm0.1415$ &	$0.2399\pm0.0603$ &	$0.9405\pm0.1414$ \\
        MolCLR~\citep{wang2022molecular} & $0.9912\pm0.2467$ &	$1.1507\pm0.1683$ &	$0.9399\pm0.1705$ & 	$1.2649\pm0.1353$ &	$0.9165\pm0.2820$ & $0.7072\pm0.1362$ & $0.2321\pm0.0847$	& $0.9574\pm0.1655$ \\
        Mole-BERT~\citep{xia2023mole} & $0.8624\pm0.0732$ & 	$1.2938\pm0.3615$ &	$1.0529\pm0.1536$ &	$1.4348\pm0.1398$ &	$1.0440\pm0.2569$ &	$0.9377\pm0.1873$ &	$0.5190\pm0.0645$	&$1.4542\pm0.3456$ \\
        ECFP~\citep{rogers2010extended} & $1.1409\pm0.2281$ &	$1.3721\pm0.0703$	& $1.2718\pm0.1531$ &	$1.5446\pm0.1133$ &	$1.2553\pm0.0985$ &	$1.0439\pm0.1244$ & 	$1.4682\pm0.2422$ &	$1.1965\pm0.1850$ \\
        GraphMVP~\citep{liu2021pre} & $1.8865\pm0.1303$ & 	$2.8140\pm0.3649$ & 	$3.0341\pm0.6418$	 & $2.7829\pm0.1195$	& $3.3965\pm0.3437$	& $3.0817\pm0.2720$ &	$3.5723\pm0.4337$	& $3.0083\pm0.2802$ \\
        NaFM& $\mathbf{0.8243\pm0.1960}$ & $\mathbf{1.1227\pm0.1604}$ & $\mathbf{0.9239\pm0.1721}$ & $1.0802\pm0.1506$	& $\mathbf{0.6927\pm0.2828}$	& $\mathbf{0.6922\pm0.1326}$	& $\mathbf{0.2265\pm0.0921}$	& $\mathbf{0.9286\pm0.1342}$ \\
        \bottomrule
    \end{tabular}
    }
    \end{threeparttable}
    \begin{flushleft}
    \small \textit{Note}: Results are reported as mean ± standard deviation. Bold values indicate the best performance in each column.
    \end{flushleft}
\end{table*}

\section{Figure legends}

\textbf{Fig. 1 Scaffold-aware pre-training framework of NaFM.}
\textbf{(a)} Overview of NaFM pre-training. After the natural product molecule undergoes scaffold-aware masking, both the masked and unmasked information are simultaneously input into a multi-layer Graph Isomorphism Network. The masked information is then processed through a projection head to predict the masked atom attributes, bond attributes, and topological information. Meanwhile, the masked and unmasked information passes through a pooling function and is used for scaffold-aware contrastive learning. \textbf{(b)} Details of scaffold-subgraph reconstruction. First, a subgraph is randomly selected from the scaffold, consisting of multiple atoms and chemical bonds. In the subgraph, both node and edge attributes are masked, and all nodes within the subgraph are fully connected (i.e., expanded into a fully connected graph), thereby masking the topological information. During the reconstruction process, the model needs to predict both node and edge attributes, while also distinguishing between real edges and those artificially added virtual edges.   
\textbf{(c)} Details of scaffold-aware contrastive learning. Different colors represent varying weights defined by scaffold similarity. \textbf{(d)} The "central dogma" of natural products. The biological source, biosynthetic gene clusters, biosynthetic pathways, and bioactivity of natural products are interconnected through the scaffold, which acts as a bridge linking these three key aspects.

\textbf{Fig. 2 Comparison of NaFM and NPClassifier across superclass categories and biosynthetic pathways.}
Bar plots show model performance in terms of AUPRC for superclass categories \textbf{(a)} and biosynthetic pathways \textbf{(b)}, including carbohydrates, amino acids and peptides, alkaloids, terpenoids, shikimates and phenylpropanoids, polyketides and fatty acids. Gray segments indicate shared performance (minimum AUPRC), whereas light blue and orange segments represent the additional AUPRC achieved by NaFM and NPClassifier, respectively. 
NaFM outperforms NPClassifier across all biosynthetic pathways and most superclass categories, with improvements of up to 24\% for amino acids and peptides. At the superclass level, NaFM achieves higher AUPRC in 53 of 71 classes, with particularly strong gains in carotenoids, alkaloids and peptide-related categories.

\textbf{Fig. 3 Representation learning enables biological source discrimination and reverse BGC functional prediction.}
\textbf{(a)} \textbf{Natural Products Atlas after representation learning.}
Compounds from the Natural Products Atlas were embedded using the pre-trained and fine-tuned molecular representations and projected into two dimensions. The learned representation organizes chemical space into distinct clusters corresponding to biological origins (animal, bacterial, chromistan, fungal and plant), revealing clear source-specific structural signatures. Representative scaffolds are highlighted. A steroid-like scaffold in the lower right illustrates a failure case, as it occurs across multiple sources and leads to ambiguous placement. The lower-left panel shows the Atlas visualization trained from scratch without pre-training, demonstrating reduced structural discrimination.
\textbf{(b)} \textbf{AUROC gap for predicting BGC-encoded protein families.}
AUROC values are shown for the 128 most frequent protein families in bacteria and fungi. Color indicates the magnitude of the AUROC gap (red to blue; deeper colors represent larger absolute differences). Across most protein families, NaFM achieves higher AUROC than baseline representations, with average improvements of ~10 points over ECFP and 20–40 points over MolCLR, particularly for families with conserved functional domains such as HATPase\_c.

\textbf{Fig. 4 Computational validation and benchmarking of NaFM-predicted inhibitors.}
\textbf{(a)}Binding free energies ($\Delta G$) of top 16 predicted compounds (blue), positive controls (orange: galantamine, tacrine), and negative control (light green: memantine). $\Delta G$ was estimated via MMPBSA (GB model) using 100 snapshots from the final 10 ns of equilibrated trajectories. Bar heights represent the mean of $n=2$ independent MD simulations (individual data points overlaid), performed as methodological repeats to ensure stability.
\textbf{(b)} Visualization of MD-simulated interactions. In the ligand, carbon atoms are shown in bright yellow and nitrogen atoms in blue. Surrounding amino acid residues are colored green. The model-selected molecule forms three hydrogen bonds with TYR337, ASP74, and SER125 (red dashed lines), and exhibits additional close-contact interactions with other residues such as TRP86 and PHE338.
\textbf{(c)} The first and second columns display several high-scoring molecules predicted by the model, along with their most similar experimentally validated actives encountered during the fine-tuning process. The third and fourth columns illustrate the maximum common substructures (MCS) between each pair of molecules and their corresponding Tanimoto similarities computed using ECFP4 fingerprints.
\textbf{(d)} Enrichment factors for NaFM and Glide across seven targets. Results represent mean ± s.d. from five-fold cross-validation ($n=5$ independent splits). Bar heights and error bars denote mean and standard deviation, respectively. Higher values indicate superior screening performance.








\end{document}